# I2C Management Based on IPbus

Shiyu Luo, Junfeng Yang, Kezhu Song, Hongwei Yu, Tengfei Chen, Tianbo Xu, Cheng Tang

*Abstract*—CBM (Compressed Baryonic Matter) is mainly used to study QCD phase diagram of strong interactions in high and moderate temperature region. Before the next generation GBTx based CBM DAQ system is built up, the DPB (Data Processing Board) layer is used in data readout and data pre-processing, where a general FPGA FMC carrier board named AFCK is used. This paper mainly describes the management of the Inter-integrated Circuit (I2C) devices on AFCK and the FMCs it carries via IPBus, an FPGA-based slow control bus used in CBM DAQ system.

On AFCK, the connection of IPBus depends on the correct initialization of a set of I2C devices, including the I2C-bus multiplexer (choosing correct I2C bus), the clock crosspoint switch (providing the 125MHz needed by 1000BASE-X/SGMII), the serial EEPROM with a EUI-48 address (providing the AFCK MAC address). An independent initial module can execute an I2C command sequence stored in a ROM, through which the FPGA can write to/read from the I2C devices without IPBus, so that the related I2C devices are correctly initialized and the necessary preparation for the IPBus start-up is fulfilled. After the initialization, a Wishbone I2C master core is used as an IPbus slave and all other I2C devices can be configured directly via IPBus. All the design has been fully tested in the CBM DPB design.

## I. INTRODUCTION

The Compressed Baryonic Matter (CBM) experiment is one of the experiments at the Facility for Antiproton and Ion Research (FAIR) in Darmstadt. The CBM experiment are composed of kinds of detectors, including Micro Vertex Detector (MVD), Silicon Tracking System (STS), Muon Chamber (MUCH), Ring Imaging Cherenkov Detector (RICH), Transition Radiation Detector (TRD),Time of Flight Detector (TOF), and Projectile Spectator Detector (PSD). All these detectors are equipped with the appropriate Front End Electronics (FEE) boards and Readout Boards (ROB) which are located near to the detectors in the irradiated area.

Before the next generation GBTx based CBM DAQ system is built up, the data from the FEE and ROB are injected into a separate intermediate layer named DPB layer, where more complex functionalities, such as slow control, time synchronization, data readout, data pre-process and data format converting, are implemented. In current CBM DAQ architecture, all these functions will be implemented in a Field Programmable Gate Array (FPGA). Considering the optimal balance between the cost and performance, a versatile FPGA-based platform, the AMC FMC Carrier Kintex (AFCK) is used in kinds of the DPB projects to support different detectors.

AFCK is a versatile prototype for high-speed control and data processing applications, which supports the extension boards connected via two HPC FMC connectors. On AFCK, a set of I2C devices can be configured from the Module Management Controller (MMC) software or from the FPGA, including:

(a) I2C bus multiplexer: Select different I2C buses controlled by FPGA or MMC;

(b) Clock crosspoint switch: Configure the interconnection of the clocks in AFCK;

(c) Programmable Clock generator: Generate local clock, with White Rabbit (WR) function supported;

(d) I2C serial EEPROM with a EUI-48 address: Store the unique board ID, which can be used as Ethernet MAC address;

(e) System Monitor ICs: Monitor the system status with temperature, voltage and current sensors;

(f) Other I2C devices, especially the I2C devices on FMCs.

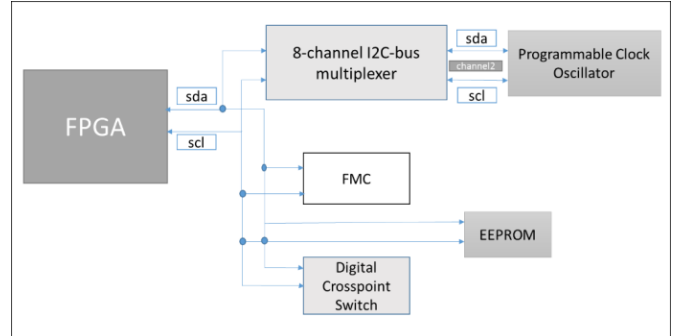

Fig. 1. I2C connection diagram

Although these I2C devices can be configured via MMC when power up, it is important for DAQ system that all these I2C devices can be on-line configured via the slow control bus. In CBM DAQ, the slow control is based on IPBus ([2]), an open-source FPGA core which controls a Wishbone-like bus via Ethernet with fully dedicated software package (c++ or Python). But the settings of IPBus depends on the correct configuration of related I2C devices, including the I2C-bus multiplexer (choosing correct I2C bus), the clock crosspoint switch (providing the 125MHz needed by 1000BASE-X/SGMII), the serial EEPROM with a EUI-48 address (providing the AFCK MAC address), which must be configured before the IPBus can be used. Hence a stand-alone, FPGA-based, programmable I2C configuration module is needed to implement the basic I2C initialization when power-up. After the IPBus link is setup, an IPBus I2C slave will fulfill the control needs for all the I2C devices connected. All these two parts consist the I2C management module for CBM DPB layers.

## II. IPBUS_I2C_CTRL MODULE DIAGRAM

The full architecture of the I2C management module can be shown as below.

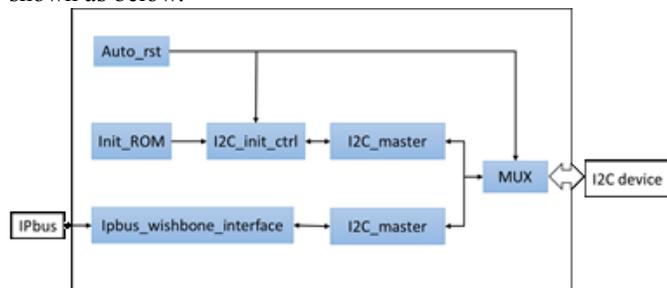

Fig. 2. I2C connection diagram

As shown in Figure 2, there are two I2C controller in the module, one is for I2C device initialization, the other is for the in-system operation on I2C device via IPbus. There is an automatical reset submodule, in which the reset signal will be asserted when power-up. This reset signal will trigger the I2C device initialization module, in which a set of I2C configuration command stored in a ROM can be read out one by one and be sent to the appointed I2C buses.

After the I2C device initialization is finished, the I2C controller switches to the IPbus-based I2C controller automatically.

The I2C controller in ipbus_i2c_ctrl core is an I2C master with WISHBONE interface, so it can be easily either accessed directly or connected to IPbus. A set of use-defined i2C configuration commands are used in the I2C device initialization. These commands should be saved in a ROM and be read out one by one by the I2C device initialization module. At the same time, the i2C device initialization module supports the input and output port for complex I2C configuration.

## III. I2C_INIT_CTRL MODULE

The initialization module is used to initialize I2C device. The initialized control module reads out the commands stored in the ROM and used to controls the I2C device. Each command stored in the ROM is 16-bit. The initialization control module reads out the command by a finite state machine (FSM). The structure of the command stored in the ROM is shown in the figure below. The 15th to 13th bits indicate which i2c bus to be selected. The 12th bit controls the read or write of the i2c bus. When the level is low, the read operation is performed. When the level is high, the write operation is performed. The 11th bit to the 8th bit represent the register port. The corresponding functions of the register are shown in Table 1. The remaining 8 bits are written to the I2C control register data.

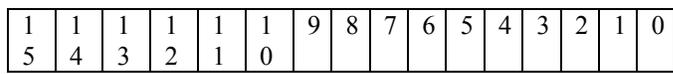

Fig. 3. Command format

TABLE I
FUNCTION OF INITIALIZING MODULE REGISTER

| Register port | I2C master register | Register port | Private register |
|---|---|---|---|
| 0000 | PRERlo(R/W) | 1000 | CR(W)/SR(R) |
| 0001 | PRERhi(R/W) | 1001 | Wait read end |
| 0010 | Control(R/W) | 1010 | Transfer read data to output port |
| 0011 | TXR(W)/RXR(R) | 1011 | Write TXR register from input port |
| 0100 | CR(W)/SR(R) | 1111 | Config done |

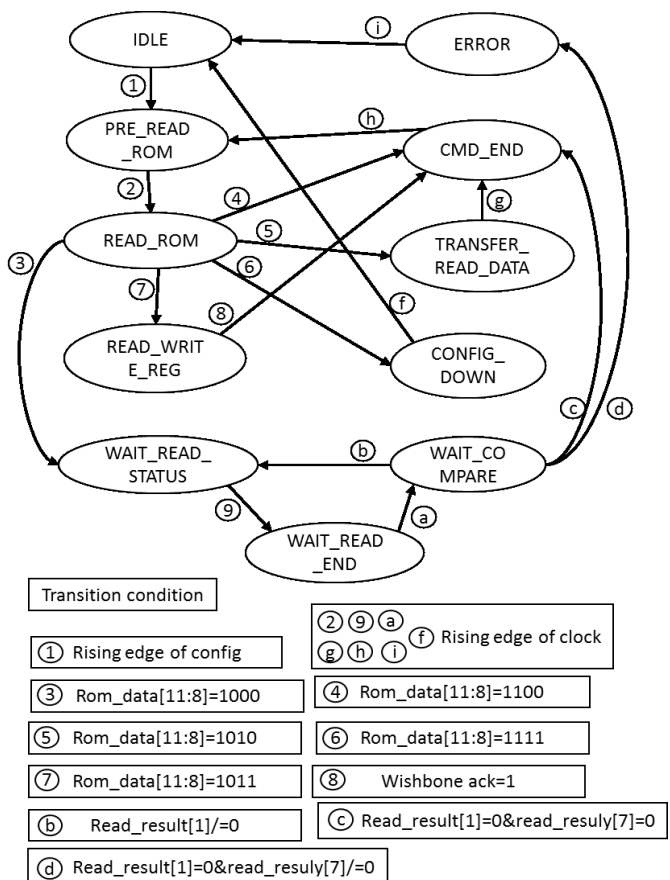

Fig. 4. Initialize module state transition diagram

The initialization controller module is controlled by a state machine. The state machine consists of a total of 11 states. The command is read in the initialization module ROM. When the 11th bit of the instruction is logic 0, the state of the state machine will be transferred from READ_ROM to READ_WRITE_REG, and then to CMD_END. According to the register port part of the command, the corresponding register in the I2C core is read or written. The state transition diagram is shown in Figure 2. When the 11th bit of the command is logical one, different operations are performed according to the register port portion of the command. The specific operations are shown in Table 1. The initialization module has an external output port and an input port. The data which is read by the I2C device can be sent to the output port and the data from the input port can be sent to the i2c device.

## IV. I2C_MASTER MODULE

After the initialization module completes the initialization, i2c master with WISHBONE interface directly connected to IPbus. The I2c master has four 8-bit wide registers, which are control register, transmit/receive register, command/status register and two clock prescale register. When IPBUS is connected to the I2C master, IPbus controls i2c device by reading or writing these registers. The control register is used to enable the I2C core. The transmit/receive register is used to store the transmitted and received data, and the command/status register is used to store the command and status information. The prescale register is used to prescal the SCL clock line. The I2C controller reads or writes these registers through the WISHBONE interface.

The initialization module directly controls the i2c device through the WISHBONE bus after power up. When the I2C device is required to read or write, the initialization module firstly writes the address of the I2C device to the transmit register. And then the I2C device is strobe. Secondly the address of the device's register is written to the transmit register. When the initialization module wants to read the strobe register, it only needs to send a read command and wait for the data to be transmitted to the I2C master through the I2C bus. Then the received data can be read from the receive register through the WISHBONE bus. Then the reading of the I2C device is done. When the register of the strobed device needs to be written, the data only needs to be written into the transmission register, then waiting for the response signal to arrive, and the data is transmitted.

## V. SOFTWARE AND ROM COMMAND FILE PROGRAMMING

Before initialization, the programmable crystal oscillator is not connected to the FPGA. So the initialization module needs a separate clock to drive it. After initialization, the system clock will connect to FPGA, and the IPBUS control module is driven by the system clock.

The initialization module reads command from the ROM to control the I2C device. The initialization module first configures the SCL clock line so that the I2C bus can transmit data normally. Secondly the initialization control module enables the I2C bus through the control register. Then the initialization module controls the I2C master to strobe the clock crosspoint switch, which is mainly used to configure the interconnection of the clocks. The initialization module controls 125MHz clock from the crystal oscillator to input to the FPGA's system clock input port and to the clock input ports of the four high-speed serial transceivers through the Clock crosspoint switch. When the high-speed serial transceiver has a reference clock, data can be received and transmitted through the fiber channel. Because the fiber channel is connected to the high-speed serial transceiver. Then initialize the control module through the I2C master strobe EEPROM. The initialization module reads the MAC address from the EEPROM and sends the MAC address to the IPBUS module through the output port of the initialization control module.

When the high-speed serial transceiver is connected to the fiber channel, the I2C aster can be directly controlled by the IPBUS. After the initialization module is completed, the I2C controller switches to the IPbus-based I2C controller. The IPbus-based I2C controller module uses Ethernet with fully dedicated software package to control WISHBONE bus. And I2C master is directly connected to WISHBONE bus. So that we can control the I2C device by software programming directly. The module is used to configure Programmable Crystal, I2C bus multiplexer, clock crosspoint switch and programmable crystal on the FMC. Firstly the module needs to initialize the I2C master. The module can set the operating frequency of the SCL and enable the I2C bus. Then the IPbus-based I2C controller module selects the I2C bus multiplexer. The programmable crystal on the AFCK carrier board is strobed through the I2C bus multiplexer. And the module would set its frequency to 156.25MHz. Then the clock from the programmable crystal oscillator is connected to the clock input port of the high-speed serial transceiver through the clock crosspoint switch, and the high-speed serial transceiver implements data transmission according to the reference clock.

## VI. CONCLUSION

The initialization control module can complete the configuration of some I2C devices. However, some I2C devices need to be configured in real time, so it is necessary to use IPbus to configure I2C devices online. The I2C device can establish connection with IPbus only after the initialization is completed. Therefore, the configuration of the I2C device on the AFCK board can be completed only when two configuration modules are used together.